PLOS COMPUTATIONAL BIOLOGY

RESEARCH ARTICLE

# Same but not alike: Structure, flexibility and energetics of domains in multi-domain proteins are influenced by the presence of other domains

Sneha Vishwanath[1], Alexandre G. de Brevern[2,3,4,5], Narayanaswamy Srinivasan[1]*

1 Molecular Biophysics Unit, Indian Institute of Science, Bangalore, India, 2 INSERM, U 1134, DSIMB, Paris, France, 3 Univ. Paris Diderot, Sorbonne Paris Cité, Univ de la Réunion, Univ des Antilles, UMR_S 1134, Paris, France, 4 Institut National de la Transfusion Sanguine (INTS), Paris, France, 5 Laboratoire d' Excellence GR-Ex, Paris, France

* ns@iisc.ac.in

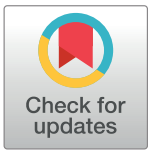

**Data Availability Statement:** All relevant data are within the paper and its Supporting Information files.

**Funding:** IISc-DBT partnership programme, India DST, India (Mathematical Biology Initiative & J.C. Bose National Fellowship, FIST program) UGC, India – Centre for Advanced Studies, Ministry of Human Resource Development, India Ministry of Research (France), University of Paris Diderot, Sorbonne Paris Cité, National Institute for Blood Transfusion (INTS, France), Institute for Health and

## Abstract

The majority of the proteins encoded in the genomes of eukaryotes contain more than one domain. Reasons for high prevalence of multi-domain proteins in various organisms have been attributed to higher stability and functional and folding advantages over single-domain proteins. Despite these advantages, many proteins are composed of only one domain while their homologous domains are part of multi-domain proteins. In the study presented here, differences in the properties of protein domains in single-domain and multi-domain systems and their influence on functions are discussed. We studied 20 pairs of identical protein domains, which were crystallized in two forms (**a**) tethered to other proteins domains and (**b**) tethered to fewer protein domains than (a) or not tethered to any protein domain. Results suggest that tethering of domains in multi-domain proteins influences the structural, dynamic and energetic properties of the constituent protein domains. 50% of the protein domain pairs show significant structural deviations while 90% of the protein domain pairs show differences in dynamics and 12% of the residues show differences in the energetics. To gain further insights on the influence of tethering on the function of the domains, 4 pairs of homologous protein domains, where one of them is a full-length single-domain protein and the other protein domain is a part of a multi-domain protein, were studied. Analyses showed that identical and structurally equivalent functional residues show differential dynamics in homologous protein domains; though comparable dynamics between *in-silico* generated chimera protein and multi-domain proteins were observed. From these observations, the differences observed in the functions of homologous proteins could be attributed to the presence of tethered domain. Overall, we conclude that tethered domains in multi-domain proteins not only provide stability or folding advantages but also influence pathways resulting in differences in function or regulatory properties.

Medical Research (INSERM, France), Laboratory of Excellence GR-Ex, reference ANR-11-LABX-0051. The labex GR-Ex is funded by the program Investissements d'avenir of the French National Research Agency, reference ANR-11-IDEX-0005-02. Indo-French Centre for the Promotion of Advanced Research/CEFIPRA for a collaborative grant (number 5302-2). The funders had no role in study design, data collection and analysis, decision to publish, or preparation of the manuscript.

**Competing interests:** The authors have declared that no competing interests exist.

## Author summary

High prevalence of multi-domain proteins in proteomes has been attributed to higher stability and functional and folding advantages of the multi-domain proteins. Influence of tethering of domains on the overall properties of proteins has been well studied but its influence on the properties of the constituent domains is largely unaddressed. Here, we investigate the influence of tethering of domains in multi-domain proteins on the structural, dynamics and energetics properties of the constituent domains and its implications on the functions of proteins. To this end, comparative analyses were carried out for identical protein domains crystallized in tethered and untethered forms. Also, comparative analyses of single-domain proteins and their homologous multi-domain proteins were performed. The analyses suggest that tethering influences the structural, dynamic and energetic properties of constituent protein domains. Our observations hint at regulation of protein domains by tethered domains in multi-domain systems, which may manifest at the differential function observed between single-domain and homologous multi-domain proteins.

## Introduction

A large proportion of proteins, coded in the genomes of diverse organisms, is constituted of more than one domain [1, 2]. Multi-domain proteins have evolved from single-domain proteins through many duplication and adaptive events [3]. Duplication and shuffling of domains have led to the emergence of various unique and novel functions using an existing repertoire of domains [3–5]. Presence of multiple domains in proteins has been reported to confer structural stability [6] and folding and functional advantages [7]. Proteins can be decomposed into domains based on various criteria namely sequence, structure, function, evolution and mobility [8, 9]. At the sequence level, domains are defined on the basis of conservation of residues over significant length; structural domains are defined on the basis of globularity and compactness; functional domains are modules in proteins which can function independently of other modules in the protein; evolutionary domains are protein modules propagating through evolution by recombination, transposition, shuffling etc. and protein modules with high correlated mobility are identified as domains according to the mobility definition [8]. It is important to note that a given protein may have different but equally valid domain annotations depending upon the basis of domain annotation [9].

Often domains in multi-domain proteins interact with one another. The role of domain-domain interfaces has been implicated in long-range allostery regulation [10–12], the emergence of a new function [13], the regulated mobility of the proteins [14] etc. In comparison to protein-protein interfaces, geometrical and chemical properties of domain-domain interfaces have been observed to be intermediate to interfaces in permanent and transient protein-protein complexes [15]. Domain interface size and linker length have been observed to influence the folding and stability of domains in multi-domain proteins [16]. The physiochemical nature of the domain-domain interface [15], the associated energetic of domain-domain interface [6] and its influence on folding in multi-domain proteins [16, 17] is well described. A recent review covers extensively the effect of domain tethering on the thermodynamics of the protein and its influence on the protein stability and folding [18]. But how protein domains behave in multi-domain proteins in comparison to single-domain proteins, has largely been unexplored and unaddressed, except some studies on the influence of tethering on the folding pathway [16, 17, 19, 20]. In the current study, we have explored how protein domains behave in multi-domain systems in comparison to single-domain systems.

For this, identical protein domains crystallized in two forms (**a**) tethered to other protein domains and (**b**) tethered to fewer protein domains than (a) or not tethered to any protein domain were studied. For example, full-length rat DNA polymerase β consists of three domains (DNA polymerase β N-terminal; DNA polymerase β and DNA polymerase β catalytic). Crystal structures are available for full-length protein (PDB id: 1BPD) and the two C-terminal domains (PDB id: 1RPL) (**Fig 1**). For the study, we have compared the properties of the second and third domains in the two crystal forms. This comparison allowed us to study the influence of the first domain on the second as well as the third domains. Further comparative dynamics analyses of homologous protein domains were carried out to understand the functional relevance of tethering of domains. Analyses reveal an intricate coupling between the domains in multi-domain systems leading to alteration in dynamics in 18 protein pairs. Structural and energetics differences were observed in half the numbers of cases studied. Differential dynamics were observed for identical and structurally equivalent functional residues of the homologous protein domain pairs. Our observations strongly suggest that tethering of domains in multi-domain proteins changes the properties of constituent domains, thus regulating the function of the entire protein.

## Results

### Tethering influences the conformation of the constituent domains

Differences in the conformation of domains were observed in comparative structural analyses of identical protein domain pairs crystallized in two forms (**a**) tethered to other protein domains (henceforth referred as **MD**) and (**b**) tethered to fewer protein domains than (a) or not

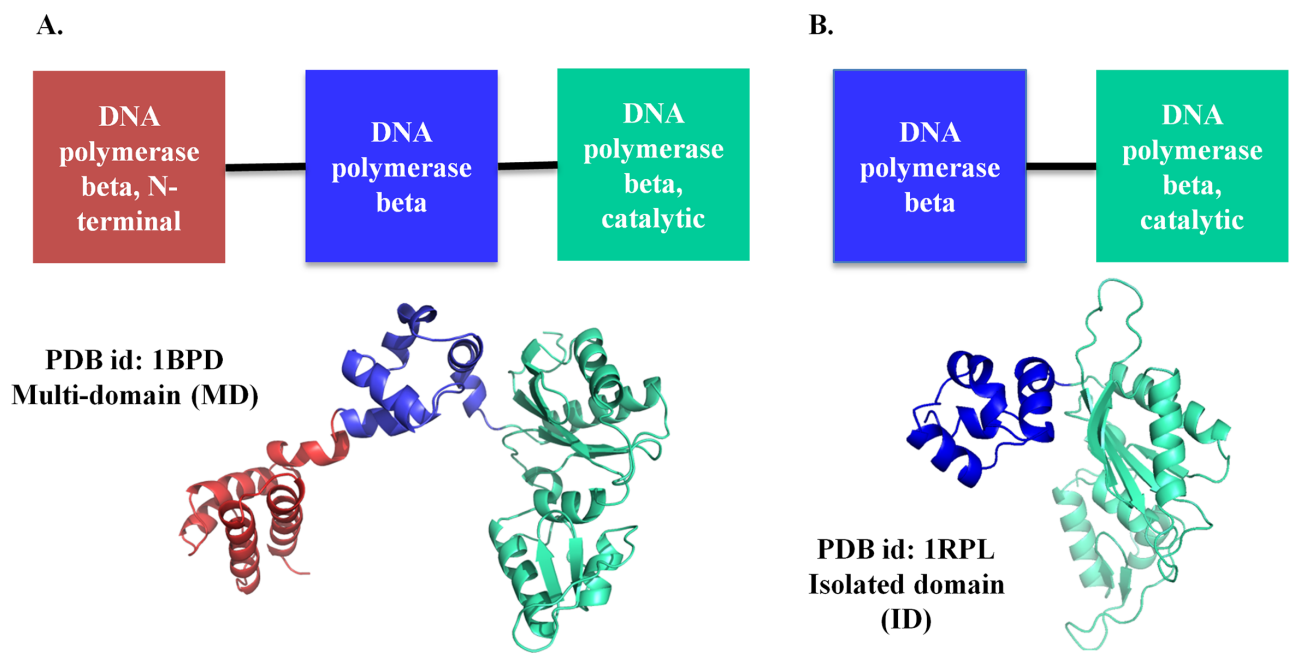

**Fig 1. An example of a domain pair used in the analysis.** Rat DNA polymerase β consists of three domains (represented in red, blue and green). The protein has been crystallized (**A**) as full length, with all the three domains (PDB id. 1BPD) and (**B**) with two C-terminal domains (PDB id. 1RPL). For all the analyses, various properties of the common domains between the members of the pair namely the DNA polymerase β domain (colored blue) and DNA polymerase β catalytic domain (colored green) are compared.

https://doi.org/10.1371/journal.pcbi.1006008.g001

tethered to any protein domain (henceforth referred as **ID**). Distributions of RMSD and GDT values for the 20 protein domain pairs are shown in Fig 2A. To delineate the differences arising due to differences in crystal packing, RMSD and GDT distributions of the protein domain

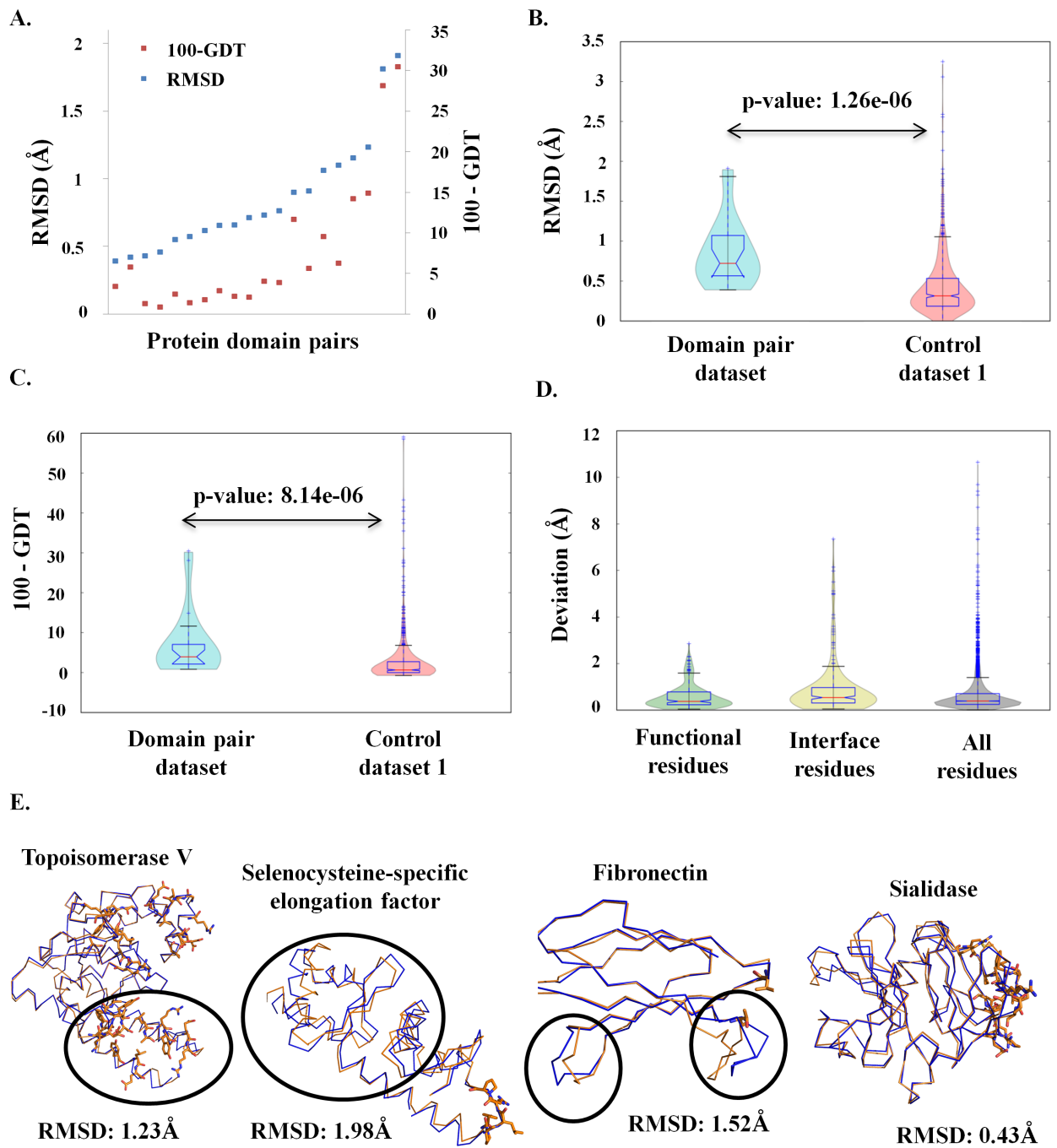

**Fig 2. Structural differences observed in the protein domain pairs.** (**A**) RMSD and GDT distributions of the 20 protein domain pairs. The X-axis represents each domain pair in the dataset and the Y-axis represents the RMSD (left) and 100-GDT (right). (**B**) RMSD distribution of the protein domain pairs (colored cyan) and the control dataset 1 (colored pink). The distributions are significantly different, two-sample KS test; p-value: 1.26e-06. (**C**) 100-GDT distribution of the protein domain pairs (colored cyan) and the control dataset 1 (colored pink). The distributions are significantly different, two-sample KS test; p-value: 8.14e-06. (**D**) Distribution of the deviations observed for functional residues, interface residues and all the residues in the dataset. (**E**) Representative examples are shown for the types of local structural deviation observed in the dataset. The domain-domain interface regions are represented in sticks and the regions showing significant structural deviation are encircled in black.

https://doi.org/10.1371/journal.pcbi.1006008.g002

pairs were compared with the control dataset 1. The control dataset 1 consists of pairs of identical monomeric proteins. Distributions of structural deviation of the protein domain pairs and control dataset 1 were observed to be significantly different (two-sample KS test, p-value: 1.26e-06 (RMSD), p-value: 8.14e-06 (100-GDT), **Fig 2B and 2C**). This suggests that structural deviations observed in the protein domain pairs are likely to be due to tethering of domains and not due to crystallization artefact. The upper quartile limits of RMSD (RMSD > 1Å) and GDT (100-GDT > 5) distributions of the control dataset 1 were taken as a cut-off to identify the protein domain pairs with significantly different conformations. RMSD and GDT distributions of the protein domain pairs suggest subtle changes in global conformation of the common protein domains for 10 cases (100-GDT ≤ 5) while 10 cases show substantial changes in the conformation (100-GDT > 5) (**Fig 2A**).

Since GDT and RMSD give an estimation of structural deviation over the entire length of a protein domain, significant structural deviations at local short stretches can be missed out. All the protein domain pairs were analyzed to identify stretches of residues showing significant structural deviation (refer structural analysis section in materials and methods). Four categories of pairs were observed: (i) only domain-domain interface showed significant structural deviation, (ii) regions other than the domain-domain interface showed structural deviation but no structural deviations were observed at the domain-domain interface, (iii) structural deviations were observed both at domain-domain interface and regions other than the domain-domain interface and, (iv) no significant structural deviation was observed between the protein domain pairs. Representative examples of the 4 case types are shown in **Fig 2E**. 9 out of the 20 protein domain pairs showed structural changes at regions other the domain-domain interface (**S1A Fig**). Further analysis of the regions with significant structural deviation shows ~14% of such regions harbors functional residues while ~24% harbors domain-domain interface residues. Functional significance of ~62% of the residues cannot be commented upon (**Fig 2D**). It has to be noted that structural deviations were observed independently of the number of domain-domain interface residues. For example, despite no interaction between the domains in fibronectin, structural deviations are observed (**S1B Fig**). The observations suggest that tethering of domains can alter the conformation of the constituent domains, with many residues apart from domain-domain interface residues showing significant structural deviation.

## Tethering of domains alters the residue-residue communication network

Previous analyses by del Sol *et al.* have shown that network property, namely residue centrality of hemoglobin and NtrC differ between the inactive and the active state of the proteins [21]. Residue centrality measures the importance of the residue in maintaining the residue-residue communication network within the protein structure. Domains in conjunction with other domains can be treated as one of the states of the protein domain and the domains in the absence of tethered domains can be treated as another state of the protein domains. Hence, a network approach was undertaken to understand the differences in residue-residue contacts, if any, for the 20 protein domain pairs. To represent residue-residue communication numerically, a network parameter namely communicability centrality (henceforth referred as coc) is used. High communicability centrality measure of a residue implies its importance in residue-residue communication in the structure. The distribution of the coc score of ID is observed to be significantly different from MD (two-sample KS-test, p-value < 2.2e-16) (**Fig 3A**). Interface residues also show differences in the coc score between MD and ID (two-sample KS-test, p-value: 1.05e-08) (**Fig 3B**). Since interface residues form intensive contacts at the domain-domain interface in MD, we expected the coc scores to be lower for interface residues in ID than MD, but ~ 30% of the residues show higher coc in ID than MD (**Fig 3B**). This observation suggests that rewiring of intra-domain residue-residue contacts of interface residues results on tethering of domains. Distribution of coc scores of non-interface residues is also observed to be different between MD and ID (two-sample KS-test, p-value < 2.2e-16) (**Fig 3C**), implying that on tethering of domains in a multi-domain system, many residues which are not part of interface region also undergo changes in the residue-residue contacts. The functional residues did not show a significant difference in the coc distribution (**S2A Fig**, two-sample KS-test, p-value: 0.04). It has to noted that ~7% (291 residues out of 4284 residues) of residues show significant differences in centrality score (|centrality score (MD)–centrality score (ID)| > 1.5) (**Fig 3A**). These 291 residues belong to 12 domain pairs in the dataset. Only ~3% of these 291 residues form a part of domain-domain interface regions. Many residues showing a significant difference in coc score (|centrality score (MD)–centrality score (ID)| > 1.5) showed low structural deviation (**S2B Fig**) implying that rewiring of the residue-residue contact can happen without any significant structural deviation. An example of the coc distribution of a domain pair (fibronectin) is shown in **Fig 3D**. Fibronectin domain shows differences in centrality score both at the domain-domain interface residues (boxed as black in **Fig 3D**) as well as residues other than domain-domain interface residues (boxed as red in **Fig 3D**).

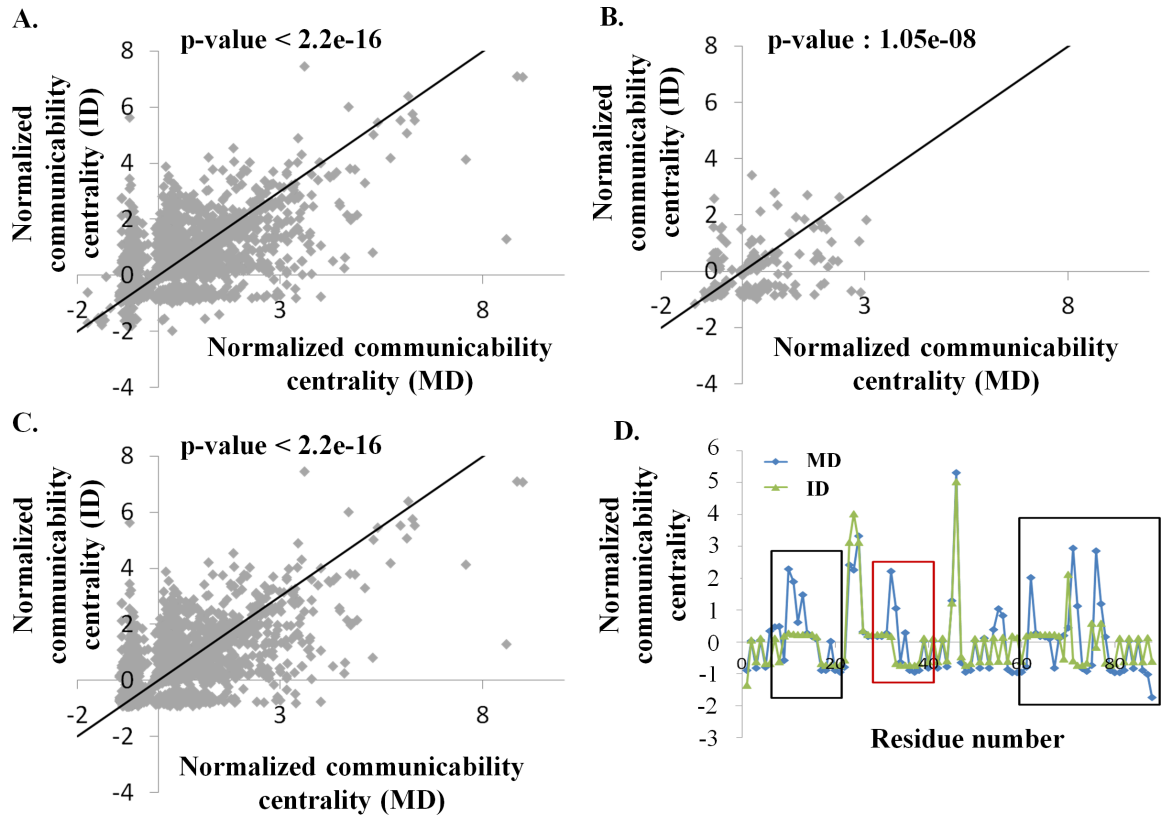

**Fig 3. Normalized communicability centrality (coc) score distribution.** Normalized coc distribution of **(A)** all the residues, **(B)** the interface residues and, **(C)** the non-interface residues of 20 protein domain pairs. The X-axis represents the normalized coc of MD and the Y-axis represents the normalized coc of ID. The solid line in the plots represents the unity line. **(D)** Normalized coc distribution of fibronectin (FNIII 10 domain). The red box encloses regions away from domain-domain interface region and the black boxes indicate the region around domain-domain interface. The X-axis represents the residue numbers and the Y-axis represents the normalized coc.

## Flexibility and coupling of fluctuations of residues change on tethering

Normal mode analysis was used to study the extent of influence of tethering on the dynamics of the constituent domains. Normal modes, accounting for 80% variance of the protein motion, were calculated for each MD and ID of the 20 protein domain pairs. To compare the flexibility of MD and ID normalized summed square fluctuation values were compared. The flexibility profiles were observed to be statistically different for all the domain pairs, except two (Fig 4A and 4B, two-sample KS test, p-value < 2.2e-16). To ensure that the differences are not an artefact of crystal packing, flexibility profiles of ID and MD were compared with two control datasets namely control dataset 2 and control dataset 3 respectively. The control dataset 2 was generated by *in silico* removal of the tethered domains from MD. The domains in the control dataset 2 (referred to as **AD**) are essentially identical to ID in sequence as well as length. The flexibility profiles of ID and AD were observed to be similar (S3A Fig). The control dataset 3 was generated by *in silico* ligation of the ID with the tethered domain of MD. This was achieved by superimposing the ID onto MD, followed by *in silico* removal of the common domain from MD and then ligation of the remaining domains of MD with ID. The multi-domains in the control dataset 3 (referred to as **swapped domain**) are essentially identical to MD in sequence and length. The flexibility profiles of MD and swapped domains were also observed to be similar (S3B Fig). The similarity of the flexibility profiles of the protein domain pairs and the control datasets ensured that the differences observed in the flexibility profiles of MD and ID are a consequence of the tethering of the domains in multi-domain systems than a crystallization artefact.

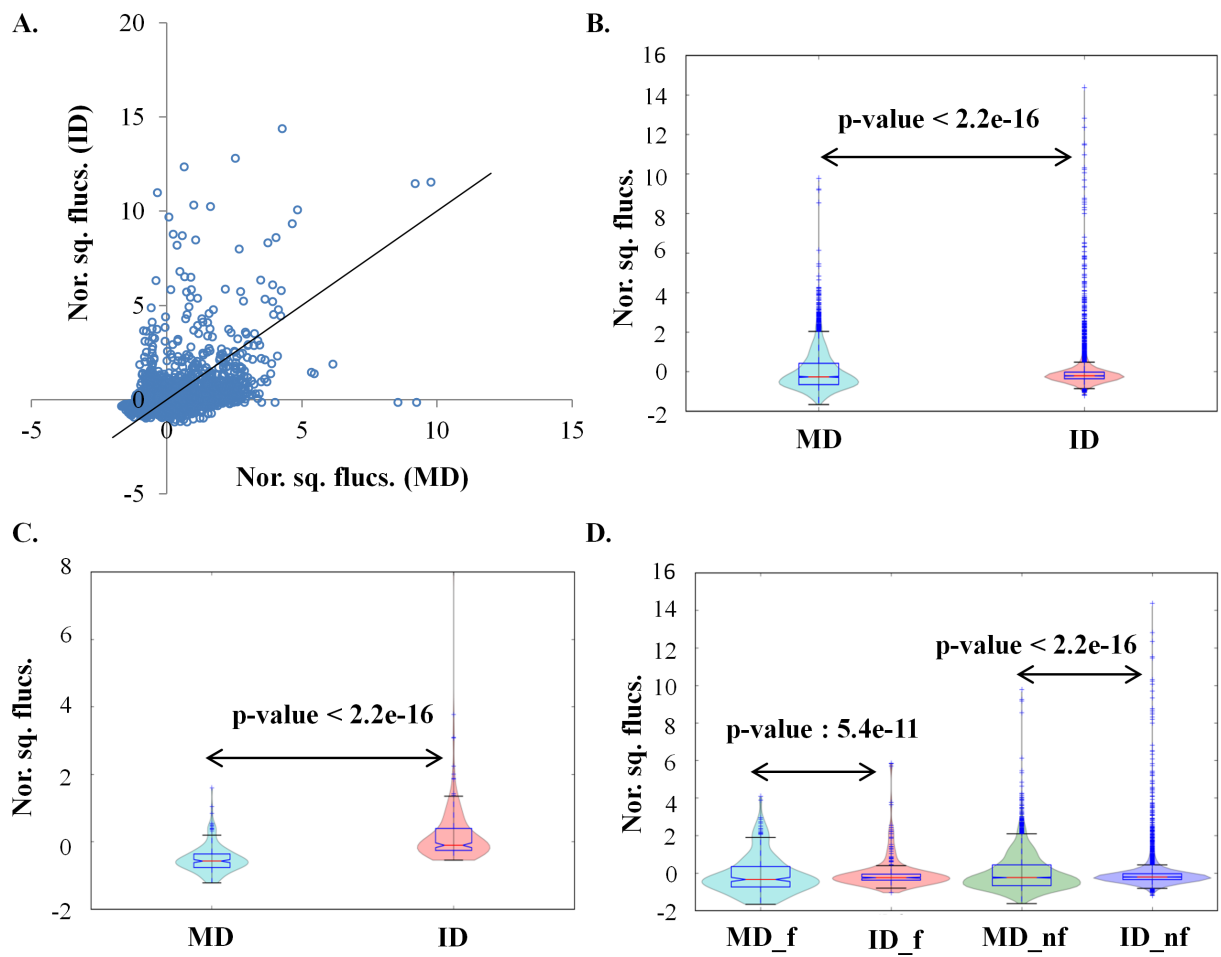

**Fig 4. Normalized square fluctuation (nor. sq. flucs.) distribution.** (**A**) Scatter plot of the normalized square fluctuation of all residues. The X-axis represents the normalized square fluctuation for MD and the Y-axis represents the normalized square fluctuation for ID. The solid line in the plot represents the unity line. Violin plots for (**B**) all residues (two-sample KS test, p-value < 2.2e-16), (**C**) interface residues (two-sample KS test, p-value < 2.2e-16) and, (D) functional residues (MD_f, ID_f, two-sample KS test, p-value: 5.4e-11) and non-functional residues (MD_nf, ID_nf, two-sample KS test, p-value < 2.2e-16).

The flexibility of the residues was observed to be different in MD and ID (Fig 4A). ~32% of the residues show higher flexibility in ID than in MD, while ~22% of the residues have higher flexibility in MD than in ID. The rest of the residues have comparable flexibilities. Higher variance in the distribution of flexibility of residues is observed for MD than ID (Fig 4B). The higher variance of the flexibility of residues in MD implies that many residues in MD show higher/lower flexibility than the mean flexibilty. To ascertain further, how the flexibility profiles of interface residues and functional residues differ in MD and ID, the flexibility distribution of the interface residues and functional residues were compared. The interface residues generally show higher flexibility in ID than MD (Fig 4C). A majority of interface residues (~70%) have higher flexibility in ID than in MD. But interestingly, ~30% of interface residues have comparable flexibility in MD and ID. Thus, some of the interface residues retain their rigidity in the isolated state as well. ~36% of the functional residues have higher flexibility in ID than MD while ~18% have higher flexibility in MD than ID (Fig 4D). Hence many functional residues are rigid in MD than ID. Many residues which are neither part of interface nor functional residues show differences in the flexibility profile (Fig 4D and S3C Fig). To ascertain whether the residues showing differences in fluctuation in MD and ID show structural deviation as well, we calculated the correlation between the two. A poor correlation (Spearman correlation coefficient: 0.25, S3D Fig) was observed between the differences in fluctuation and structural deviation, suggesting tethering of domains can alter the dynamic properties of protein domain without significant structural conformation change.

Residue-residue communication in protein domains is important for the function and structural integrity of proteins. Residues can relay information to other residues either by forming contacts or through synchronization of dynamics. To understand the influence of tethered domain on the synchronization of dynamics of residues in protein domain, the extent of correlation of fluctuation among residues (henceforth referred as cross-correlation) was studied. Higher number of residues with high cross-correlation value ($|$cross-correlation$| \geq$ 0.7) was observed for MD (~22%) as compared to ID (~10%) (Fig 5A). This observation implies that residues show tight coupling ($|$cross-correlation$| \geq$ 0.7) in the case of MD but no or weak coupling in the ID ($|$cross-correlation$| <$ 0.7). Moreover, clusters of high correlation were observed in the case of MD; which often corresponded to sub-domains or domains or super-secondary structures in the spatial coordinate. The matrices of MD and ID were observed to have a low similarity (low $R_v$ coefficient) for all the domain pairs except two (Fig 5B). A representative example (fibronectin) is shown in Fig 5C and cross-correlation matrices for 20 protein domain pairs are shown in S4 Fig. To ensure that differences are not observed due to crystal packing or other artefact, $R_v$ coefficient between cross-correlation of ID and control dataset 2 and cross-correlation between MD and control dataset 2 were calculated (S5 Fig). The comparison ruled out any other factor apart from tethering for the behavior observed. An important point to note here is that this characteristic has been observed irrespective of the number of interactions between the domains. For example, the domains in rat DNA polymerase β do not interact with each other but still, low $R_v$ coefficient is observed (**1BPD in Fig 5B**).

Molecular dynamic studies were carried out for 3 domain pairs from the dataset to study the synchronization of motions in the domain at all-atom level. These 3 pairs of domains were selected based on the number of interfacial residues between the domains. Tight coupling of motions was observed not only between the C-alpha of residues but also between the side-chains of residues in MD (S6 Fig). While weak or no coupling was observed for side-chains of residues in ID. Thus, molecular dynamics analysis for 3 pairs showed that higher cross-correlation between residues in MD is manifested not only at the backbone level, as observed also from NMA, but also at the side-chain level. All the observations imply that tethering of domains in multi-domain proteins alters the flexibility as well as the synchronization of the fluctuations of residues of the constituent domains.

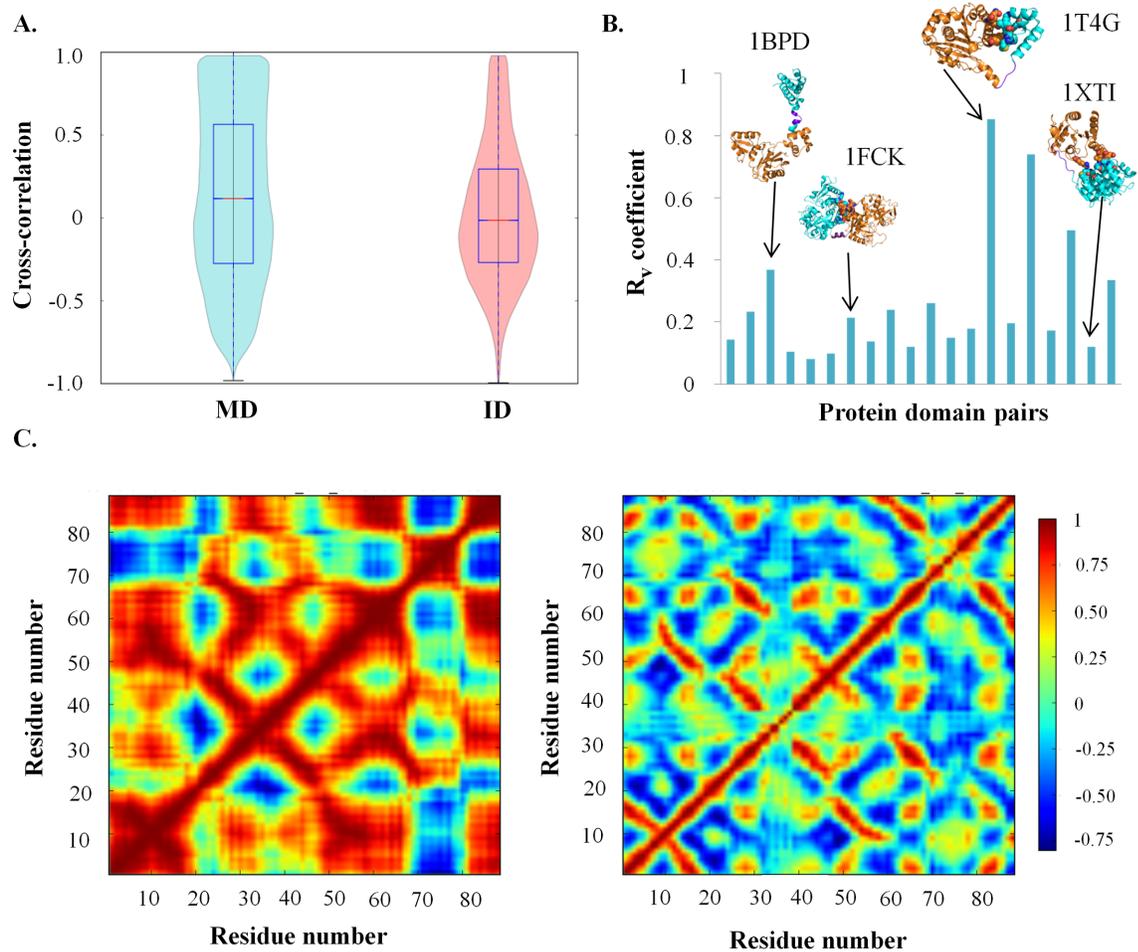

**Fig 5. Cross-correlation analysis of protein domain pairs.** (A) Distribution of the cross-correlation values for 20 protein domain pairs. A higher number of residues show high positive cross-correlation values in MD as compared to ID. (B) Distribution of $R_v$ coefficient for 20 protein domain pairs. Cartoon representation of few proteins along with the PDB id is shown above the corresponding $R_v$ coefficient. The spheres represent the interface residues. (C) Cross-correlation matrices of fibronectin (FNIII 10 domain). The left panel is for MD and the right panel is for ID. The color bar represents the cross-correlation values.

## The stability of residues and residue-residue contacts changes on tethering

From the network analysis of the structure of the 20 protein domain pairs, it was observed that certain residues show significant differences in the communicability centrality score. We further wanted to study whether this rearrangement in the intra-domain residue-residue contacts, as represented by communicability centrality score, changes the energetic stability of the residues and residue-residue contacts. Frustratometer algorithm [22] was used to study the effect of tethering on energetics distribution of residues. The algorithm calculates a parameter, single residue level frustration (SRLF), for each residue in the structure. Two parameters, configurational frustration index and mutational frustration index, are calculated for all the contact pairs in the structure. SRLF measures the energetic stability of the residue with respect to every other amino acid at that position. Configurational frustration index measures the stability of the contact pair with respect to every other configuration the contact pair can take during the folding process. Mutational frustration index measures the stability of the contact pair with respect to every other amino acid combination at that position. Mathematically, frustration index is the Z-score of the energy of the native with respect to the decoys. A residue or a contact is considered as minimally frustrated if the frustration index is greater than 0.78, highly frustrated if the frustration index is less than -1 and neutrally frustrated if frustration index is in between -1 and 0.78 [22].

The frustration indices were calculated for the 20 protein domain pairs. Though the distribution of SRLF of MD and ID were observed to be largely comparable (two-sample KS test, p-value: 0.98) (Fig 6A) but ~12% (region II, III, IV, VI, VII and VIII of Fig 6A) of the residues showed differences in the single residue level frustration (SRLF) with 5 residues (region III and region VII of Fig 6A) showing drastic substitution from high frustration to minimal frustration and vice-versa. These 12% residues are distributed over the entire domain dataset i.e. each domain pair have at least one residue showing different frustration indices. Residues apart from domain-domain interface residues and functional residues were also observed to differ in the frustration index (S7A Fig). Moreover, differences in the frustration index of MD and ID were observed to be independent of the structural deviation observed. Equivalent numbers of substitutions were observed at structural deviation greater than 1Å and lower than or equal to 1Å (Fig 6D). Similar trends as that of SRLF were observed for configurational frustration and mutational frustration (Fig 6B, 6C, 6E and 6F) but a higher number of contacts showed differences in configurational frustration type as compared to mutational frustration type. Many residues which are neither domain-domain interface residues nor associated with function showed differences in the frustration type of contact (S7B and S7C Fig). The differences suggest that when a protein domain tethers to another domain not only the stability of entire domain [6] or the folding rates differ as reported earlier [16] but the stability of the residues as well contact pairs changes for few cases. Since a larger number of contacts were observed to be configurationally frustrated (higher the configurational frustration index; more stable the conformation during the folding process) in comparison to mutationally frustrated, it hints that the domains may sample different conformations during the folding process in MD and ID, as have been reported earlier in literature for some multi-domain proteins [16–20].

## Domain tethering influences the functions of homologous protein domains with different domain compositions

To understand further the influence of tethering of domains on the function of proteins, a comparative analysis was performed for homologous domain pairs, where one member is a single-domain protein while the other member is a part of a multi-domain protein. Both the members are full-length gene products. Four pairs of proteins namely (**a**) phosphoribosylanthranilate isomerase from *E. coli* (PDB id: 1PII) and *Jonesia denitrificans* (PDB id: 4WUI), (**b**) cyclophilin from *Bos taurus* (PDB id: 1IHG) and *Homo sapiens* (PDB id: 3ICH), (**c**) sialidase from *Micromonospora viridificaciens* (PDB id: 1EUT) and *Homo sapiens* (PDB id: 1SO7) and, (**d**) hexokinase-1 from *Homo sapiens* (PDB id: 1HKC) and *Saccharomyces cerevisiae* (PDB id:

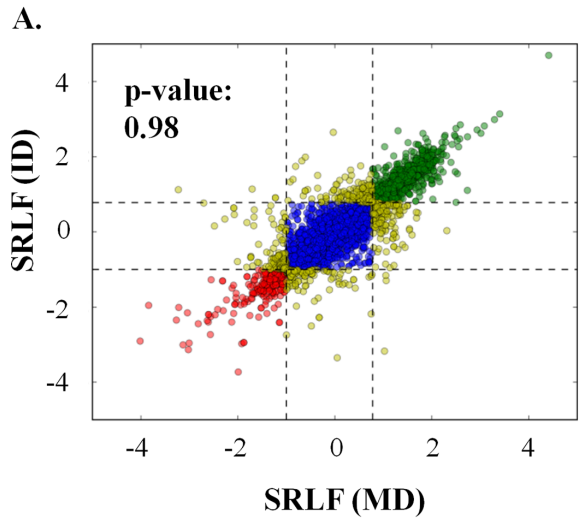

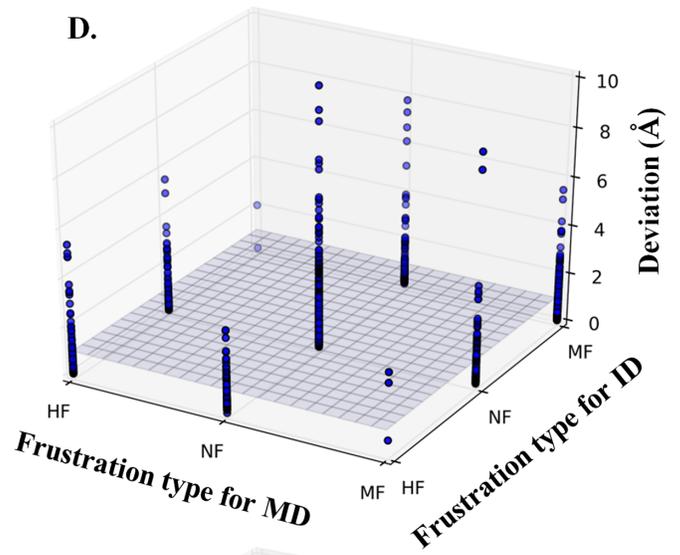

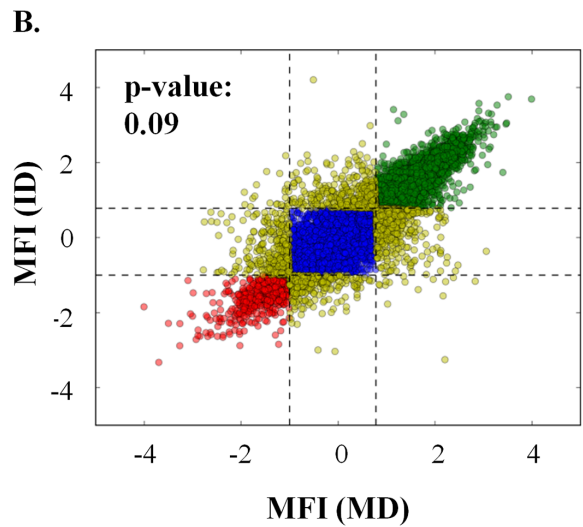

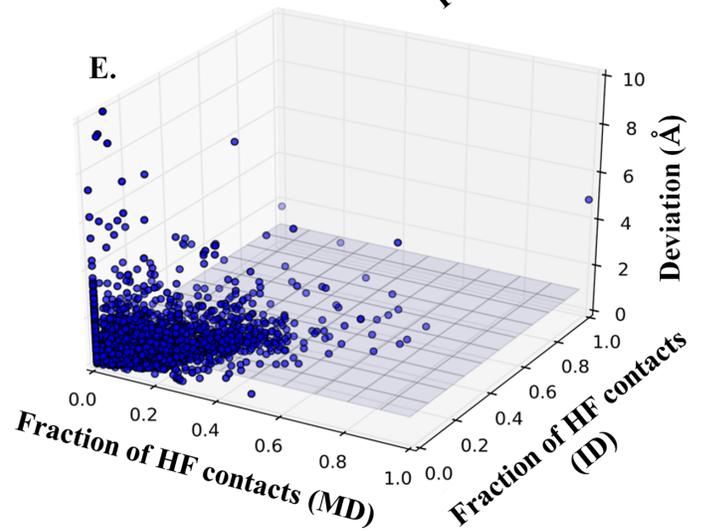

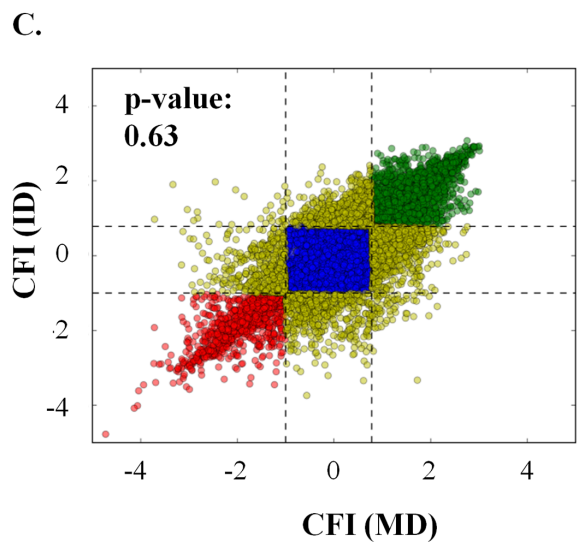

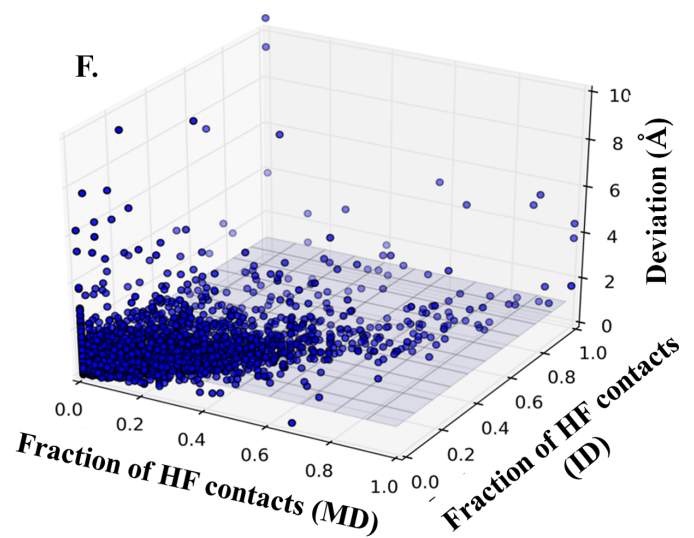

**Fig 6. Frustration index (FI) distribution.** Scatter plot of the distribution of **(A)** single residue level frustration (SRLF) (two-sample KS test, p-value: 0.98), **(B)** mutational frustration index (MFI) (two-sample KS-test, p-value: 0.09) and, **(C)** configurational frustration index (CFI) (two-sample KS test, p-value: 0.63) of residues for all the domain pairs. The X-axis represents the frustration index for MD and the Y-axis represents the frustration index for ID. Residues which are minimally frustrated (MF) in both MD and ID are represented as green filled circles, residues which are neutrally frustrated (NF) in both MD and ID are represented as blue filled circles and residues which are highly frustrated (HF) in both MD and ID are represented as red filled circles. The residues showing differences in the type of frustration between MD and ID are represented as yellow filled circles. The dotted lines represent the cut-off used for the definition of frustration index (MF: FI ≥ 0.78; HF: FI ≤ -1 and NF: -1 < FI < 0.78). 3-D plot of the distribution of **(D)** frustration type, **(E)** fraction of highly mutationally frustrated contacts of each residue and, **(F)** fraction of highly configurationally frustrated contacts of each residue and the structural deviation observed between the corresponding C-alphas on superposition. A plane (grey color) is drawn at the deviation value of 1Å.

3B8A) were studied. The four domain pairs have sequence identity in the range of 27–56% with RMSDs in the range of 1.3–2.2Å (Fig 7). Since the homologous proteins differ in their amino acid sequences, only the dynamic properties of the protein were compared. The dynamics of the proteins were studied using normal mode analysis. For the comparative analysis, *in-silico* multi-domain chimeras of the single-domain proteins were generated. This was achieved by superposing the single-domain protein on the multi-domain protein, followed by *in-silico* removal of the homologous domain from the multi-domain protein and ligation of the domains. This *in-silico* protein will henceforth be referred as a chimera. For the hexokinase-1 protein, since the two-functional domains show gene duplication, the chimera was generated by superposing the single-domain on both the domains of multi-domain. Thus the two halves of the chimera of hexokinase-1 are identical. The flexibility and the cross-correlation coefficient of the functional residues were compared between single-domain proteins, multi-domain proteins and the chimeras for understanding the influence of tethering of domains on the function of proteins. Only topologically equivalent and identical functional residues of the homologous domain pairs were compared to minimize the influence of nature of residues.

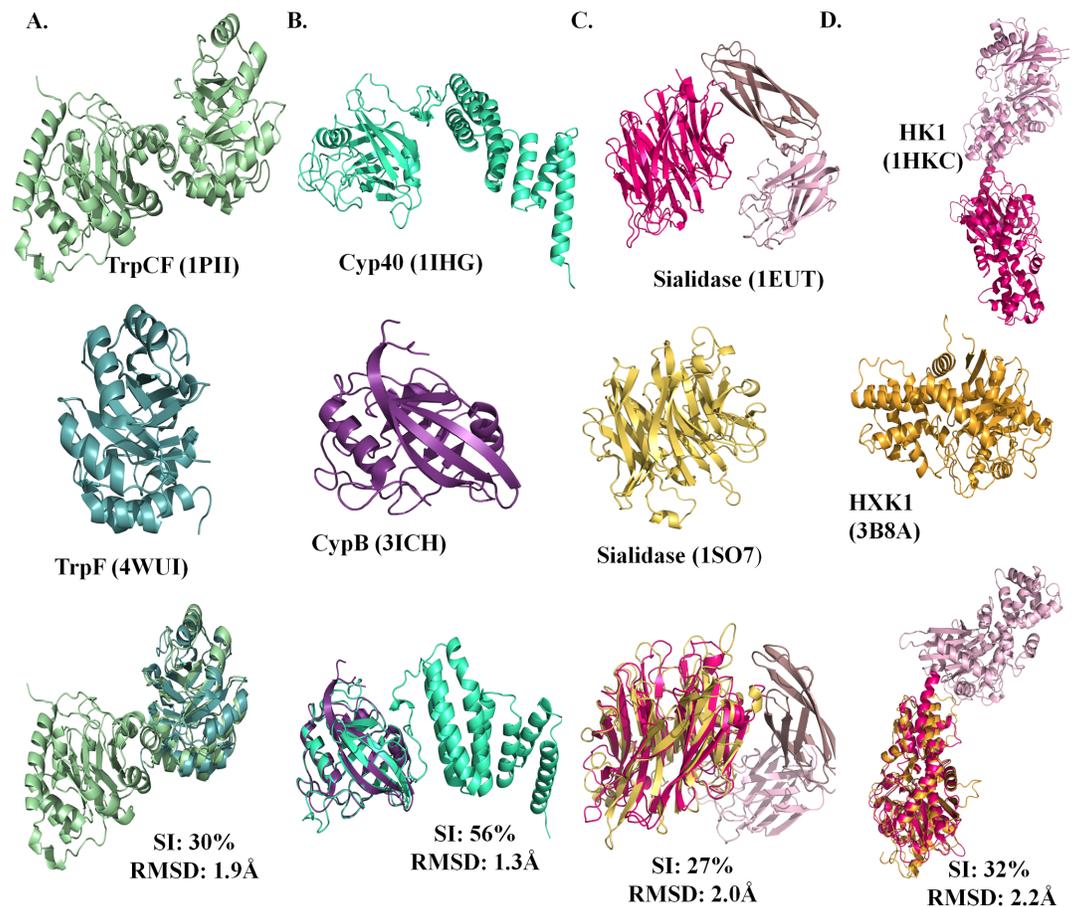

**Fig 7. Homologous domain pairs.** Four homologous domain pairs namely **(A)** phosphoribosylanthranilate isomerase from *E. coli* and *Jonesia denitrificans*, **(B)** cyclophilin from *Bos taurus* and *Homo sapiens*, **(C)** sialidase from *Micromonospora viridifaciens* and *Homo sapiens* and, **(D)** hexokinase-1 from *Homo sapiens* and *Saccharomyces cerevisiae*. The top panel shows the multi-domain protein, the centre panel shows the single-domain protein and the last panel shows the superposition of homologous domains, with sequence identity (SI) and RMSD values.

Normalized square fluctuations of functional residues were compared between the single-domain and multi-domain proteins. The functional residues have lower flexibility (normalized square fluctuation < 0) in both single-domain and multi-domain proteins ([Fig 8](#)). Residues important for function or structural integrity are known to show lower flexibility [23]. Nonetheless, the flexibility of functional residues is lower in the multi-domain proteins as compared to the single-domain proteins ([Fig 8](#)). The flexibility of functional residues in the multi-domain protein and the chimera is observed to be similar (**S8 Fig**) except in the case of siali-dase. This observation implies that increase in the rigidity of functional residues is a conse-quence of tethering of domains in multi-domain proteins. The differences in the flexibility of the functional residues can contribute towards differences reported in the functions of homol-ogous protein domains, which are discussed later. To further understand the alteration in the dynamic properties of the domain, cross-correlation of the functional residues were studied. High correlation of motions was observed among functional residues for multi-domain pro-tein in comparison to single-domain proteins (**Fig 9, upper row**). The single-domain proteins showed weaker cross-correlation among residues for all the cases (**Fig 9, middle row**). The cross-correlation between functional residues was comparable between the multi-domain and chimera for all the cases, except hexokinase-1 (**Fig 9, lower row**). The observations suggest that alteration in the synchronization of motion is a consequence of tethering.

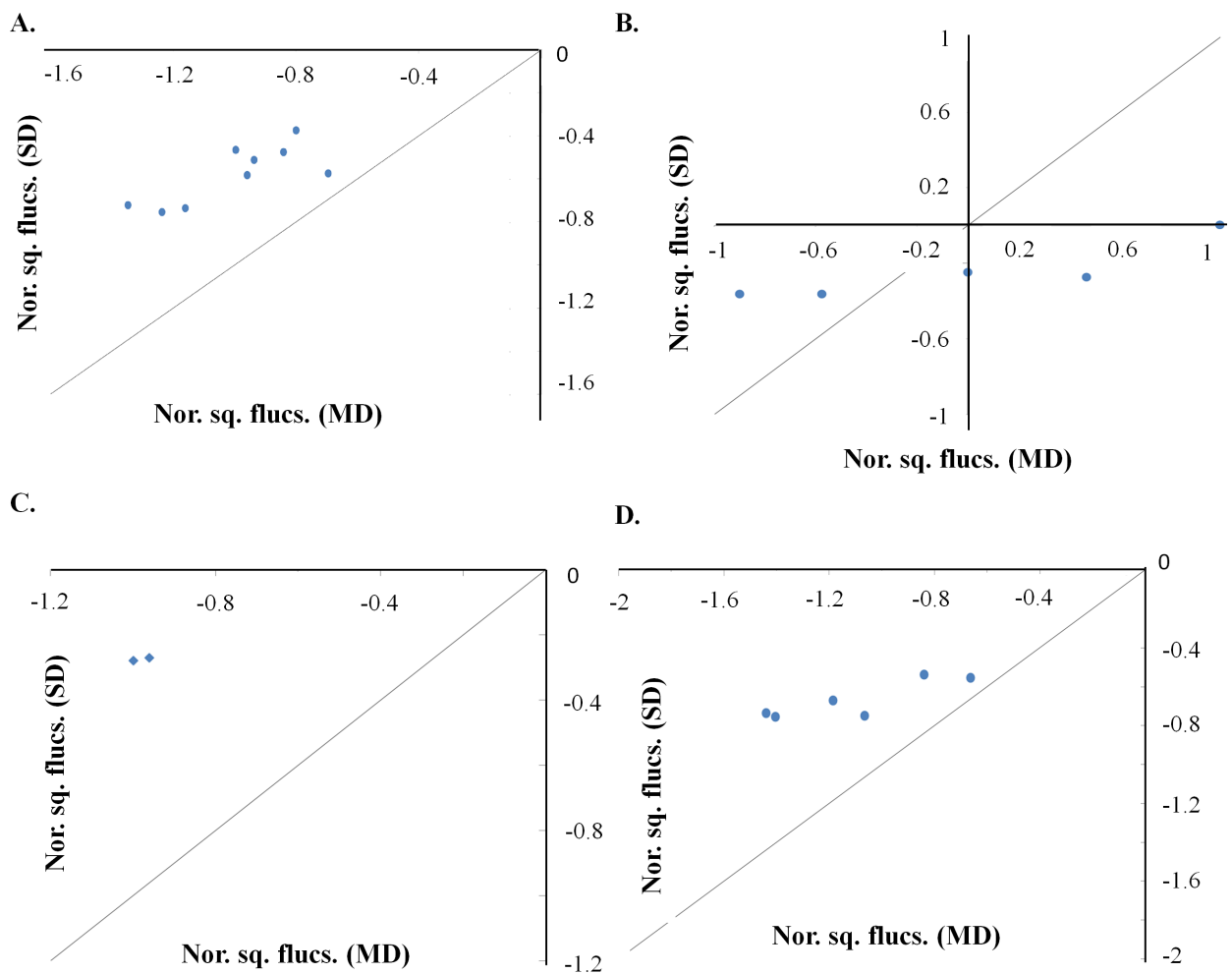

**Fig 8. Distribution of the normalized square fluctuation of the functional residues.** Scatter plots for functional residues of (**A**) phosphoribosylanthranilate isomerase, (**B**) cyclophilin, (**C**) sialidase and, (**D**) hexokinase-1. The X-axis represents the normalized square fluctuation for multi-domain protein (MD) and the Y-axis represents the normalized square fluctuation for single-domain protein (SD). The solid line in all the plots represents the unity line.

For cyclophilin, the multi-domain protein is known to be less sensitive to cyclosporin as compared to single-domain cyclophilin [24]. Detailed analysis of cyclophilin single-domain protein showed the cyclosporin binding pocket shows low cross-correlation because of the closing movement of the pocket; but the multi-domain cyclophilin is superseded by domain-domain motion, where the functional residues move in the same direction resulting in high cross-correlation values (**S9 Fig**). This differential dynamics can provide a rationale for the lower sensitivity towards cyclosporin of the multi-domain protein in comparison to single-domain protein. The closing movement of the functional residues in the single-domain pro-tein can hold the ligand better than the observed motion of the residues in the multi-domain

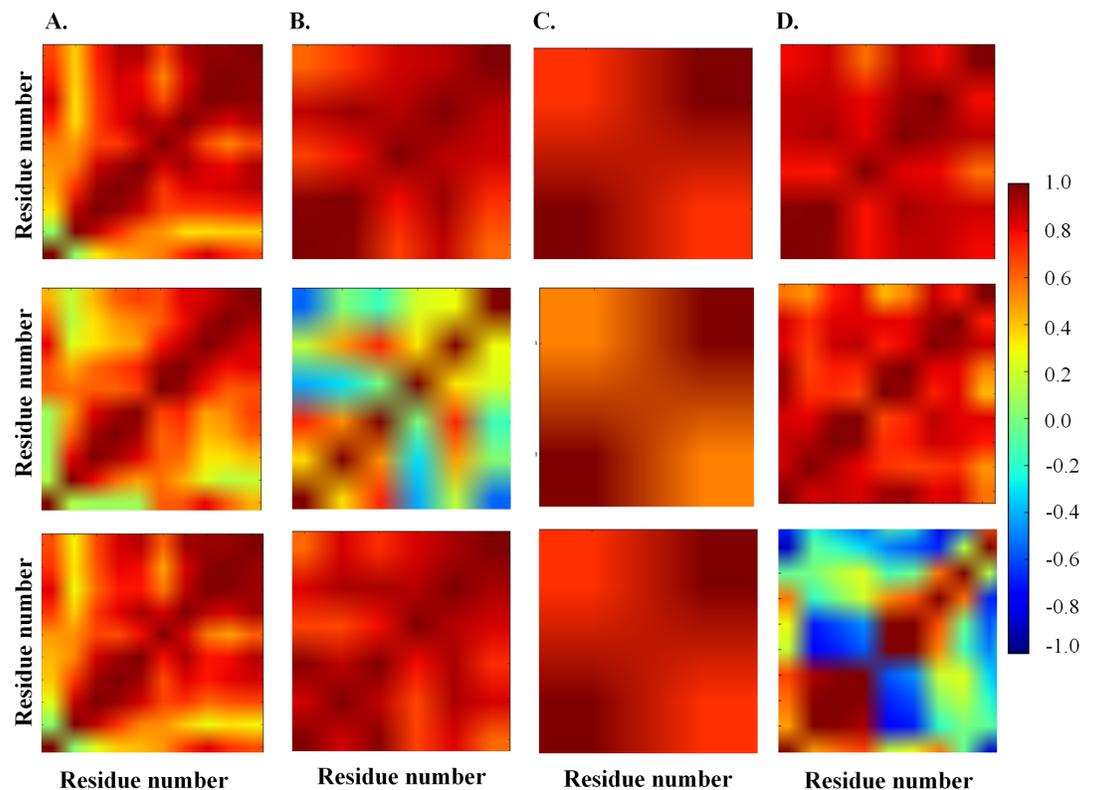

**Fig 9. Cross-correlation matrices of the functional residues.** (**A**) Phosphoribosylanthranilate isomerase, (**B**) cyclophilin, (**C**) sialidase and, (**D**) hexokinase-1. The first row represents the matrices for multi-domain proteins, the second row represents the matrices for single-domain and, the last row represents the chimeras. The X-axis and the Y-axis represent the residue number of the functional residues. The color bar represents the cross-correlation values.

https://doi.org/10.1371/journal.pcbi.1006008.g009

protein. The single-domain hexokinase-1 protein has higher $K_m$ (300 μM) [25] as compared to multi-domain protein (32 μM) [26]. The glucose-binding pocket is at the interface of sub-domains for both multi-domain and single-domain protein. The sub-domain movement in single-domain protein is superseded by the domain movement in multi-domain protein. The low-frequency global motion in multi-domain protein allows better-synchronized motion of the binding pocket as compared to single-domain protein (Fig 9D). Weaker correlation between residues in single-domain hexokinase-1 as compared to multi-domain hexokinase-1 can explain the different $K_m$, despite identical binding protein. From these analyses, we argue that tethering of domains influences the function of the constituent domains.

Chimera hexokinase-1 also exhibited an interesting feature. Though the structure and sequence of the two protein domains in the chimeric hexokinase-1 is identical, the domains exhibited different flexibility profile (S10A Fig). It has to be noted that while constructing the chimera of the yeast hexokinase-1, a stretch of 9 amino acids from the C-terminal of the first domain and a stretch of 9 amino acids from the N-terminal of the second domain were removed to relieve short contacts at the domain-domain interface region and linker region. To ensure that the differences are not observed due to this specific amino-acids deletions, the flex-ibility profile of the natural single-domain yeast hexokinase-1 (3b8a in S10B Fig) was com-pared with the *in-silico* generated model of the yeast single-domain hexokinase-1 with 9 amino acids deleted from the N-terminal (3b8a_N in S10B Fig) and the *in-silico* generated model of the yeast single-domain hexokinase-1 with 9 amino acids deleted from the C-terminal (3b8a_C in S10B Fig). The flexibility profiles were observed to be identical (S10B Fig), imply-ing that the differences in the flexibility profile are only due to the tethering of domains and not due to deletion of the amino-acids. The observations suggest that the differences observed in the constituent domains of multi-domain protein depend on the order of the domain in the multi-domain proteins. The cross-correlation between the functional residues in the N-termi-nal and C-terminal domain also differs (S10C Fig). A number of positively correlated motions were observed in the C-terminal domain than in N-terminal domain. 6 pairs of functional resi-dues viz. 173–210, 173–211, 174–210, 173–211, 176–210 and 176–211 exhibit anti-correlation motion in the N-terminal domain while the same residue pairs exhibit positively correlated motion in the C-terminal domain. We hypothesize that the differences in the nature of correla-tion of the fluctuation of the functional residues in the N-terminal and C-terminal domain may have given rise to the differential functional activity of the two domains in human hexoki-nase-1 at the first duplication event during evolution. The C-terminal of the human hexoki-nase-1 is catalytically active while N-terminal is catalytically inactive.

## Discussion

Conformational and structural alterations have been observed in proteins as they bind to other proteins [27, 28]. This line of thinking is extended in the current work to understand the struc-tural, dynamic and energetic effects of tethering of protein domains in multi-domain proteins on the constituent domains. The extent of similarity between the physical and geometrical properties of protein-protein interaction and domain-domain interaction in multi-domain proteins [15] motivated us for the study. A dataset of 20 protein domain pairs of known 3-D structure has been used in the analysis. Each pair comprises of an entry with one or more domains of a multi-domain protein and the other entry has at least one additional domain tethered. Fifty percent of the protein domain pairs show differences in the global conformation on tethering. Rewiring of some intra-domain residue-residue contacts was observed in 12 pro-tein domain pairs. Normal mode and molecular dynamics analyses of the domain pairs sug-gested that the flexibility of residues differs between domain in isolation and domain in multi-domain protein. Tight coupling of fluctuation was observed between residues in multi-domain proteins as compared to domain in isolation for all the domain pairs except one. These differences in the fluctuation and coupling of fluctuation are observed due to the shift from low-frequency local motion in isolated domain to low-frequency global motion in multi-domain systems. The stability of ~12% of residues and residue-residue contacts changed on tethering in all the domain pairs. Many of the differences in the intra-residue contacts, dynamics and energetics of the residues were observed without any significant structural deviation. These results strongly suggest that tethering of domains in multi-domain proteins influences the conformation, intra-domain residue-residue contact map, dynamics and the stability of residues and residue-residue contact of domains. Structural, dynamic and energetic differences were observed for many residues apart from domain-domain interacting residues in many domain pairs. These differences at regions spatially away from domain-domain interface could have allosteric origin; where the domain-domain interface region is the orthosteric site, the regions showing alteration are the allosteric site and the perturbation being tethering of domains. Allosteric alteration of proteins by altering the flexibility or correlated motion of the side-chains has been reported for some proteins [12, 29–32]. For example, the isolated WW domain and PPIase domain of human Pin1 protein has been shown to retain substrate binding and isomerase activity *in vitro;* but genetic studies showed that the WW domain is essential for *in vivo* Pin1 activation [12, 29, 30]. The WW domain regulates the activity of the PPIase domain by altering the flexibility and the extent of correlation of motion of side-chain of the three catalytic loops without much conformational changes [12, 29, 30].

To gain further insights on how tethering of domains influences the function of proteins, comparative dynamics analyses were carried out for 4 pairs of homologous domains, where a member in a pair is a multi-domain protein and the other member is a single-domain protein which is a homologue of one of the domains in the other protein in the pair. In each pair, only identical and structurally equivalent functional residues were analyzed. Functional residues were observed to be more rigid in all the multi-domain proteins than the single-domain proteins. This rigidity of functional residues is observed due to superseding of the low-frequency local motion of the single-domain protein by the low frequency global domain-domain motion in the multi-domain proteins. The low-frequency global domain motion alters the synchronization of residue-residue motion of functional residues in multi-domain proteins as compared to single-domain homologues. Differences in the catalytic activity reported for these homologous domain pairs can be a manifestation of these alteration in fluctuations. Combined with our observations on the identical domain pairs, it can be concluded that tethered domains in multi-domain proteins influence the function of domains by affecting the dynamics of the domains. Identical functional residues were observed to have different dynamics depending on the domain order, as exemplified by the chimera hexokinase-1 in our study. The N-terminal and the C-terminal domains of the chimera hexokinase are identical in sequence and conformation, but the flexibility and the synchronization between functional residues differ between the two domains. Similar observation was made by Kirubarkaran *et. al.* Artificial two-domain proteins were generated by fusing the natural protein domains PDZ3 and SH3 with five artificial domains. Observed differences in the fluctuation of the residues in PDZ3/SH3 domains were found to be dependent on the order of the domain construct for many cases [33]. These observations suggest that domains are not tethered during evolution at random but as a design to modulate the function of the constituent domains. Since dynamic alterations are observed in all the domain pairs; irrespective of the number of interface residues, size of the constituent domains, directionality of domain order or the fold (as defined in SCOP) of the domains (**Table A and B in S1 Text**), it can be concluded that dynamic allosteric regulation of domains is an intrinsic property of multi-domain proteins. This observation reinforces reports by others in literature that allostery is an intrinsic property of globular protein and allosteric regulation is prevalent in many multi-domain proteins [11, 34–36]. Alteration in the dynamics of the domain without any significant conformational difference by the tethered domain can be a great tool by evolution to modulate the function of same domain in different multi-domain proteins without altering the fold or structure of the domain, which otherwise can be an expensive process.

Alterations in the covalent structure of proteins such as post-translational modifications are known for causing changes in the conformation and/or nature of dynamics at the site of modification and around [37–40]. For example, phosphorylation of the activation loop of kinases such as cAMP-dependent kinase and CDK is well known to alter the conformation of the kinase extensively, enabling transition between inactive and active forms [41–43]. In our work, we considered pairs of identical domains, one in isolation and the other tethered to another domain. This pair can be viewed as though the domain in isolation is "modified" covalently in the other structure in the pair i.e. a domain and a domain linker region is covalently attached at one of N or C-terminus of the domain of interest. Clearly, this "covalent modification" in the terminus will have an influence on the structure/dynamics of the domain in the neighborhood of covalent attachment or possibly, even at a distant site. Interactions between the domain-domain linker and the flanking domains are common for all the examples studied in this work. Indeed, such interactions are present even in the examples where the direct domain-domain interactions are not present as the two domains are spatially well separated, for example cyclophilin and hexokinase-1. We believe that interactions between domain-domain linker and the domain of interest play a significant role in conferring alterations in structure, dynamics and correlated motions we observe in comparison with isolated domains. Since alterations in dynamics were observed independent of the number of amino acids in the linker (**Table A and B in S1 Text**), we believe that the effects depend on the presence of linker than the length of the linker. Role of linker residues in the allosteric communication between domains has been suggested by others as well in the literature [11, 33, 44–47]. All these observations suggest that the tethered domain and linker region can act as a scaffold for allosteric modulation of domains. The study presented here can be further exploited in designing new domain combination with desired activity.

## Materials and methods

### Dataset preparation

**Structures of same protein domains in tethered and isolated forms.** The dataset for the analyses was prepared using domain definition available in SCOPe 2.03 ver. [48]. All the protein domain entries in SCOPe 2.03, whose structures have been elucidated using X-ray crystallography, were filtered for domain entries with single chain both in the asymmetric unit (ASU) and biological unit (BU) and no other biological entity is present. This constraint was applied to ensure that no variances are observed due to the oligomeric condition of the protein or its interaction with other biomolecules. The filtered dataset was then clustered at 100% sequence identity using CD-HIT [49] and pairs of proteins were selected from each cluster, where the pair of entries differed only by the presence of extra domain/s in one entry and the common identical domains differed maximum by ± 10 amino acid residues. Further, those pairs of proteins that have different cognate ligands were removed from the dataset. This step assures that no variances are observed due to the presence of ligands. After applying all the aforementioned filters, 20 pairs of structures (**Table C in S1 Text**) were obtained. The small number of entries in the dataset is the reflection of the rigorous quality checks we have employed and the relatively small number of multi-domain proteins in the PDB and SCOPe database (21% of the dataset). It is also important to note that the longer member in each pair need not be a full-length gene product. For all the analyses, various features of the common domains of each pair have been compared. Information on function-associated residues (mentioned as functional residues) was taken from the literature survey and UNIPROT [50]. Domain boundary definition for all the multi-domain protein structures was taken from SCOPe 2.03 [48]. Domain-domain interface residues were identified using the sum of van der Waals radii + 0.5Å cut–off as inter-atomic distance criteria. Any pair of residues with the atomic distance less than the sum of van der Waals radii + 0.5Å was considered as domain-domain interface residues.

**Single and multi-domain proteins with shared homologous domains–homologous domains data set.** SCOPe 2.03 database was screened for protein entries solved using X-ray crystallography, having a single chain in both asymmetric and biological unit and not bound to any other macromolecule. Crystal structures of only full-length proteins were selected. The selected proteins were clustered using local alignment tool of CD-HIT [49] at 25% sequence identity. Each cluster was searched for pairs of homologous domains, where one entry is a single-domain protein and the other entry is a multi-domain protein. Further, these pairs of entries were refined for the presence of same cognate ligands and same conformation. It was ensured that the EC number was same for the homologous protein pairs. It enabled us to establish a functional relationship between the homologous protein pairs and also to extract the catalytic site information reliably. Further, it was made sure that the binding sites share high sequence identity (>70%) and show low structural deviation (< 2Å). 4 pairs of homologous proteins satisfied these restraints and were used for further analysis (**Table D in S1 Text**).

**Control dataset 1:** All the protein domain entries in SCOPe 2.03 [48], whose structures have been elucidated by X-ray crystallography, were filtered for domain entries with single chain both in the asymmetric unit (ASU) and biological unit (BU) and no other biological entity is present. This constraint was applied to ensure that no variances are observed due to the oligomeric condition of the protein or its interaction with other biomolecules. The filtered dataset was then clustered at 100% sequence identity using CD-HIT [49] and pairs of proteins with identical domain composition were selected. 607 pairs of protein were identified and it constituted the control dataset 1. This dataset was generated to study the impact of crystal packing on the conformation of protein domains.

**Control dataset 2:** The control dataset 2 was generated by *in-silico* removal of the uncommon domain from MD of domains pairs. The length and sequence of the domains (referred as AD) are identical to ID. This dataset was generated to study the impact of crystal packing on dynamics of proteins.

**Control dataset 3:** The control dataset 3 was generated by superimposing the ID onto MD followed by *in-silico* removal of the common domain from MD and ligation of the superimposed ID with the uncommon domain of MD. The length and sequence of the domains (referred as swapped domains) are identical to MD. This dataset was generated to study the impact of crystal packing on dynamics of proteins.

## Structural analysis

Proteins in the datasets were structurally aligned using TM-align [51]. For the same-domain dataset, the structural variations were studied at the global and local level. Global Root Mean Square Deviation (RMSD) and Global Distance Test–Total Score (GDT-TS) [52] score were used to define global deviations. GDT-TS, henceforth mentioned as GDT, is used to define structural similarity between domains of identical sequences. Unlike RMSD it is largely insensitive to outliers arising especially due to differences in loop conformations. It is defined as the number of alpha carbons falling within a distance cut-off from the corresponding Cα of the other structure. MAXCLUSTER, an improved version of the maxsub algorithm [53], with a cut-off of 4Å was used for calculation of GDT score. High GDT scores are indicative of a low structural deviation between the proteins.

For studying structural variation at the local level, regions of residues that show significant structural deviation as compared to other regions of the structure were compared. For this, the distance between corresponding Cα atoms of the protein pairs after superimposing the structure onto each other was calculated. All the residues, whose distance between corresponding Cα (s) is more than twice the standard deviation from the mean of the distance distribution of all the residues, were identified as region showing significant structural deviation. For homologous protein domain pair, only RMSD has been calculated to quantify the structural differences.

## Protein structure network analysis

To capture differences, if any, in residue-residue communication within proteins; undirected and unweighted networks of protein structures were constructed. The network was constructed for repaired structures (refer following section on dynamics). Each node in the network represents Cα and each edge represents the interaction between the nodes provided the distance between Cα atoms is less than or equal to 5Å. Network property communicability centrality was calculated using NetworkX [54] module of python. Communicability centrality quantifies the extent to which a node communicates with its neighbour. High communicability centrality measure of a residue implies its' importance in inter-residue communication in protein structure. Numerically, it is the summation of all the closed walks of all lengths starting and ending at a node.

## Dynamics: Normal mode analysis and molecular dynamics

To study dynamics of domains, we have used two approaches namely normal mode analysis (NMA) and molecular dynamics. Crystal structures were energy minimized using GROMACS package [55] with conjugate gradient as the energy minimization method. Prior to energy minimization, the structures were repaired for missing residues and missing atoms. The missing residues were modelled using Rosetta 3.4 [56] and missing atoms were built using WHAT IF 10.1a algorithm [57]. Normal modes were calculated by generating coarse-grained anisotropic network model (ANM) for proteins, with 15 Å as the cut-off for connecting the nodes. Distance-dependent spring constants (the closer the nodes, stiffer is the edge) were used for the edges. Calculation of normal modes as well as the associated calculations and analyses were done using the ProDy package [58]. For the analyses, only the normal modes contributing to 80% variations were studied and fluctuation values contributed by first five N-terminal residues and last five C-terminal residues were removed. Furthermore, correlation of fluctuation between each residue pairs, termed as cross-correlation, was compared. The similarity between cross-correlation matrices has been measured using distance independent measure called $R_v$ coefficient [59]. $R_v$ coefficient measures the closeness of a set of points represented as a matrix. It is a multivariate generalization of Pearson correlation coefficient.

Molecular dynamics was performed to study the correlation of fluctuation at the all-atom level for 3 pairs. Molecular dynamics was performed using GROMACS package. The proteins were simulated using Charmm 27 force field [60] and SPC water model [61] in a dodecahedron box. The system was energy minimized using steep descent after addition of appropriate counter ions to balance the charges. The system was appropriately equilibrated for 100 ps using V-rescale and 100 ps using Parrinello-Rahman. The final production run was performed once for 400ns.

## Frustration calculation

Energetics calculation was performed only for a dataset of identical protein domains. As the homologous protein domains differ in sequence identity, it is futile to compare their energetics. Frustratometer algorithm [22] was used to perform the energetics calculation. The algorithm systematically perturbs each residue and contact to generate the decoys and compute energy according to Associative Memory Hamiltonian with Water-mediated interaction energy function (AMW) [62]. Then the energy of the native protein is compared with the energy distribution of the decoys to calculate the frustration index, which is the Z-score of the energy of the native with respect to the decoys. A residue or a contact is considered as minimally frustrated if the frustration index is greater than 0.78, highly frustrated if the frustration index is less than -1 and neutrally frustrated if frustration index is in between -1 and 0.78 as defined in [22].

AMW is a coarse-grained energy function where the backbone is represented as Cα, O and the side chain is reduced to Cβ, the position of N and C is generated considering the ideal geometry of the peptide bond. AMW energy function consists of five non-local energy terms namely Lennard-Jones 6–12 potential, H-bond potential, compactness potential, burial potential and water-mediated interaction potential. A pair of amino acids is considered to form a contact if the inter Cα distance is less than or equal to 5Å. Each contact is perturbed either by mutating each interacting residue pair to every other amino acid pair but keeping all other interaction parameters same as the native structure. Then the effective energy of the native contact is compared with the decoys to access the energetic stability of the contact to mutation. So, it provides a qualitative measure of the energetic feasibility of mutation of such contacts. The frustration index calculated by this method is termed as mutational frustration index. Another way of perturbing the contacts is by displacing the location of each contact thus sampling the possible configurations which can be taken by the contacts during folding. The frustration index calculated in such a way is termed as configurational frustration index. Similar to contacts, each residue is perturbed to every other amino acid and other configurations to evaluate the stability of residue in the native structure to all these perturbation. The frustration index calculated by this method is termed as Single Residue Level Frustration (SRLF).

## Statistical analysis

All the statistical analyses were performed using R package.

## Supporting information

**S1 Fig. Examples of domain pairs showing significant structural deviations at regions other than domain-domain interface residues. (A)** 9 pairs out of 20 protein domain pairs show structural deviations at regions other than domain-domain interface residues. Such regions are encircled in black while the domain-domain interface residues are represented in sticks. These deviations were observed irrespective of the number of domain-domain interface residues. (**B**) Cartoon representation of fibronectin (MD is colored in bright orange and ID is colored in blue). The domains in fibronectin are observed to be non-interacting. (TIF)

**S2 Fig. Normalized communicability centrality (coc) score distribution. (A)** Normalized coc distribution for the functional residues. The X-axis represents the coc of MD and the Y-axis represents the coc of ID. The solid line in the plot represents the unity line. (**B**) Scatter plot of the absolute difference of coc of MD and ID and the structural deviation between corresponding Cα. The solid line is drawn at the absolute difference of coc of 1.5. Values above 1.5 signify the significant difference between coc of MD and ID.
(TIF)

**S3 Fig. Flexibility profile of control datasets.** (**A**) Distribution of normalized square fluctuations of ID and the control dataset 2 (AD). The two distributions are similar (two-sample KS test, p-value: 0.253). (**B**) Distribution of normalized square fluctuations of MD and the control dataset 3 (Swapped domain). The two distributions are similar (two-sample KS test, p-value: 0.813). (**C**) Scatter plot of the normalized square fluctuation of all the non-functional and non-interface residues, The X-axis represents the normalized square fluctuation for MD and the Y-axis represents the normalized square fluctuation for ID. The solid line in the plot represents the unity line. (**D**) Distribution of absolute difference in normalized square fluctuation and the structural deviation between corresponding Cα. Spearman correlation coefficient for the distribution is 0.25. The inset shows the expanded view of the dense region.
(TIF)

**S4 Fig. Cross-correlation matrices for all domain pairs.** The first and the third column represent the cross-correlation matrices for MD in domain pairs. The second and the fourth column represent the cross-correlation matrices for the corresponding ID of domain pairs. The residues have been observed to be tightly dynamically coupled in MD as compared to ID. The color bar represents the cross-correlation value. The X-axis and the Y-axis represent the residue number.
(TIF)

**S5 Fig. $R_v$ coefficient distribution.** For all the domain pairs, three $R_v$ coefficients have been plotted. $R_v$ coefficients have been calculated for cross-correlation matrices of MD and the control dataset 2 (AD), cross-correlation matrices of MD and ID, and cross-correlation matrices of ID and the control dataset 2 (AD). The high $R_v$ coefficient for ID and SD implies that the low $R_v$ coefficients observe for MD and ID are not a manifestation of crystallization artefact.
(TIF)

**S6 Fig. Molecular dynamics simulation.** For three domain pairs, molecular dynamic simulations were carried out for 400 ns. (**A**) Selenocysteine elongation factor (**B**) Fibronectin (**C**) DNA repair and recombination protein radA protein. Cartoon representations of the domain pairs are shown on the extreme left, with common domain colored blue. The left side matrices represent the cross-correlation of the residues at C-alpha level and the right matrices represent the cross-correlation at the all-atom level. The first rows in (A, B, and C) represent MD and the second rows in (A, B, and C) represent ID. The color bars represent the cross-correlation value.
(TIF)

**S7 Fig. Frustration index distribution.** Scatter plot of the distribution of (**A**) single residue level frustration (SRLF) (**B**) configurational frustration index and (**C**) mutational frustration index. The left panel represents functional residues, centre panel represents interface residues and the right panel represents non-interface and non-functional residues for all domain pairs. The X-axis represents the frustration index for MD and the Y-axis represents the frustration index for ID. Residues which are minimally frustrated (MF) in both MD and ID are represented as a green filled circle, residues which are neutrally frustrated (NF) in both MD and ID are represented as blue filled circle and residues which are highly frustrated (HF) in both MD and ID are represented as red filled circles. The residues showing the difference in the type of frustration between MD and ID are represented as a yellow filled circle. The dotted lines have been drawn to represent the cut-off used for the definition of frustration index (MF: FI $\geq 0.78$; HF: FI $\leq$ -1 and NF: -1 $<$ NF $< 0.78$).
(TIF)

**S8 Fig. Distribution of the normalized square fluctuation of the functional residues.** Scatter plots for functional residues of (**A**) phosphoribosylanthranilate isomerase, (**B**) cyclophilin, (**C**) sialidase and, (**D**) hexokinase-1. The X-axis represents the normalized square fluctuation for multi-domain protein (MD) and the Y-axis represents the normalized square fluctuation for chimera protein. The solid line in all the plots represents the unity line.
(TIF)

**S9 Fig. Vector representation of the movement of the functional residues of cyclophilin.** In the upper left panel and the upper right panel, the Cα residues are represented as blue spheres and the directionality and the magnitude of the first mode of motion for functional residues is represented by a red arrow. Longer the arrow higher is the magnitude of the motion. The upper centre panel represents the space-fill representation of the functional residues (cyclosporin binding pocket). The lower left and right panel represents the cross-correlation matrices for functional residues for multi-domain and the single-domain protein respectively. The pocket closing movement of the protein in single-domain protein is superseded by the global domain movement.
(TIF)

**S10 Fig. Flexibility and cross-correlation profile of the chimeric hexokinase-1.** (**A**) Distribution of normalized square fluctuation of the N-terminal (3b8a_art_N) and C-terminal (3b8a_art_C) of the chimeric hexokinase-1. Certain residues show differences in the flexibility between the two domains. (**B**) Distribution of normalized square fluctuation of the full-length yeast hexokinase-1 (3b8a), yeast hexokinase-1 with 9 residues removed from the N-terminal (3b8a_N) and yeast hexokinase-1 with 9 residues removed from the C-terminal (3b8a_C). The distributions are highly similar, suggesting that the amino-acids deletions do not affect the flexibility profile. (**C**) Cross-correlation plot of the functional residues of chimeric proteins. The numbering of the functional residues has been kept same for the N-terminal and the C-terminal domain. The solid line separates the X-axis and the Y-axis into the N-terminal and the C-terminal domain. The X-axis and the Y-axis represent the residue number of the functional residues.
(TIF)

**S1 Text. The text file contains information on the dataset for the same domain and homologous domain dataset and the information on the linker length, number of interface residues, SCOP fold, size of the domain and the nature of differences observed.**
**Table A.** SCOP fold annotations, number of interface residues, length of the linker, size of the domain and the nature of differences observed for the identical domain pair dataset.
**Table B.** SCOP fold annotations, number of interface residues, length of the linker, size of the domain and the nature of differences observed for the homologous domain pair dataset.
**Table C.** PDB codes and macromolecule name of pairs of same proteins used for the analysis with their respective resolution (Å). **Table D.** PDB codes of homologous domains pairs with their resolution (Å).
(DOCX)

## Acknowledgments

SV thanks, all the lab members for useful discussion.

## Author Contributions

**Conceptualization:** Alexandre G. de Brevern, Narayanaswamy Srinivasan.

**Data curation:** Sneha Vishwanath.

**Formal analysis:** Sneha Vishwanath.

**Funding acquisition:** Alexandre G. de Brevern, Narayanaswamy Srinivasan.

**Investigation:** Sneha Vishwanath, Alexandre G. de Brevern, Narayanaswamy Srinivasan.

**Methodology:** Sneha Vishwanath.

**Project administration:** Alexandre G. de Brevern, Narayanaswamy Srinivasan.

**Supervision:** Alexandre G. de Brevern, Narayanaswamy Srinivasan.

**Validation:** Sneha Vishwanath.

**Writing – original draft:** Sneha Vishwanath.

**Writing – review & editing:** Sneha Vishwanath, Alexandre G. de Brevern, Narayanaswamy Srinivasan.